\documentclass[aip,jcp,preprint,nofootinbib,endfloats]{revtex4}
\usepackage{epsfig,epsf,subfigure,amsmath,amssymb}
\usepackage{overpic}
\usepackage{graphicx}
\usepackage{comment}

\begin{document}
\author{Santosh Mogurampelly}\thanks{Corresponding author}
\email{santosh@physics.iisc.ernet.in; santoshcup6@gmail.com}
\affiliation{Centre for Condensed Matter Theory, Department of Physics, 
Indian Institute of Science, Bangalore - India 560012.}
\author{Bidisha Nandy}
\affiliation{Centre for Condensed Matter Theory, Department of Physics,
Indian Institute of Science, Bangalore - India 560012.}
\author{Roland R. Netz}
\affiliation{Fachbereich Physik, Freie Universit\"{a}t Berlin, 14195 Berlin, Germany.}
\author{Prabal K. Maiti}
\affiliation{Centre for Condensed Matter Theory, Department of Physics, 
Indian Institute of Science, Bangalore - India 560012.}

\title[]
{Elasticity of DNA and the effect of Dendrimer {Binding}}

\begin{abstract}
Negatively charged DNA can be compacted by positively charged dendrimer{s}
and the degree of compaction is a delicate balance between the strength of the
electrostatic interaction and the elasticity of DNA.
We report various elastic 
properties of short double strand{ed} DNA 
(dsDNA) and the effect of dendrimer {binding} using 
fully atomistic molecular dynamics and 
numerical simulations. In equilibrium at room
temperature, the {contour length distribution 
$P(L)$ and end-to-end distance distribution $P(R)$ are nearly 
Gaussian, the former gives an estimate of
the stretch modulus $\gamma_1$ of dsDNA in quantitative agreement with the literature value. The bend
angle distribution $P(\theta)$ of the 
dsDNA also has a Gaussian form and allows to extract a persistence length, $L_p$ of 43 nm. When the dsDNA 
is compacted by positively charged dendrimer, 
the stretch modulus stays invariant but the effective 
bending rigidity estimated from the end-to-end distance
distribution} decreases dramatically 
due to backbone charge neutralization
of dsDNA by dendrimer. We support our observations with 
numerical solutions of the worm-like-chain (WLC) 
model as well as using non-equilibrium dsDNA stretching 
simulations. These results are helpful in 
understanding the dsDNA elasticity at short length 
scales as well as how the elasticity is modulated when 
dsDNA binds to a charged object such as a dendrimer or protein.
\end{abstract} 
\maketitle
\pagebreak

\section{Introduction}
Many fundamental biological processes of life such as
DNA replication, translation and transcription involve
interaction of DNA with proteins where the elasticity
of DNA is crucial as a short segment of DNA is tightly
wound around {proteins}. The length scales involved in such
biological processes are less than the persistence length
(50 nm) of dsDNA and are of interest to study. The 
advancement of micromanipulation techniques in the last 
decades {allows} to perform manipulation experiments with single molecule DNA to 
understand its mechanical properties. Elastic properties 
of short DNA of few 10's of base-pairs play a significant 
role in many cellular processes \cite{shore,widom,rippe}. 
Extensive experimental work has been done on DNA 
elasticity in the decade \cite{bustamante1992,bustamante1994,bustamante3,perkins,marko1995}, 
but most studies involve long DNA of more than few 
hundreds of base-pairs in length. Many cellular processes 
involve unzipping of local DNA base-pairs when
proteins bind to DNA with specific interactions 
\cite{roland1999,roland2000}. 
Poly amido amine (PAMAM) dendrimers are hyperbranched 
polymers and can be considered as model proteins with 
many protein-like structural similarities 
\cite{maiti2004pamam,maiti2005pamam,svenson2005}. PAMAM dendrimers are positively 
charged at neutral and low pH \cite{maiti2008pamam} and 
can bind negatively charged DNA. Earlier we have 
studied the interaction of DNA with a dendrimer 
\cite{maiti2006pamam,nandy2011} at varying pH conditions 
and {showed that} the binding energy of the DNA-dendrimer complex 
increases with the size of the dendrimer \cite{nandy2011}.
In this paper we try to understand how the elasticity of 
DNA is altered while complexed with a dendrimer, which can 
be viewed as a model protein. The length
scale of the binding area at the DNA binding site covered by 
typical proteins spans few base-pairs, which is similar 
to the size of dendrimers.\\

The end-to-end distance ($R$) distributions ($P(R)$) of semi-flexible
polymers in the context of their elasticity have been 
studied in the past decade extensively
\cite{thirumalai1995,frey1996,samuelsinha2002,dabhi2002,
stepanow2002,winkler2003,dietler2005,ranjit_menon2005,thirumalai2006}.
Recently both experiments and simulations have focused 
on the short length scales to study the elasticity of 
short dsDNA \cite{seol2007,mathewfenn2008,yuan2008,ranjith2013}.
Mazur \cite{mazur2006,mazur2007,mazur2009,mazurpre2009} 
has studied the elastic properties of dsDNA 
using atomistic simulations based on the probability 
distributions of end-to-end distance, bending angle etc.
To study the effect of dendrimer binding, recent studies 
\cite{chen2011,dootz2011} using small angle X-ray scattering
revealed that dsDNA has different bending modes depending on
the dendrimer charge density.
{The case of a} semi-flexible
polymer interacting with an oppositely charged 
sphere {was} treated as a simple model case of
DNA wrapping around histones, forming nucleosome core
particles (NCP). This study revealed how the wrapping propensity 
is influenced by the ionic strength of the
solution \cite{roland1999,roland2000} and how structures very 
similar to chromatin appear \cite{roland2011}.
Understanding short length scale elastic behavior of dsDNA 
is important since the local bending and unzipping of dsDNA 
can occur when it binds to proteins. The length scales
over which protein binds to dsDNA in a DNA-protein complex are in the nano meter scale, which
is less than the persistence length of dsDNA. However,
models based on worm-like chain (WLC) largely fails to explain elastic behavior 
of dsDNA on such short length scales.\\ 

With the advance of single molecule experimental techniques like 
optical tweezers, magnetic tweezers and atomic force microscopy 
(AFM), it has become possible to study structural 
details of single DNA (both dsDNA and ssDNA) under 
external force at varying physiological conditions. 
Several experimental and theoretical groups have studied 
\cite{bustamante1994,marko1995,bustamante2000,gosse,lebrun,cluzel,santoshjpcm,santoshbj} 
structural transformations of DNA by external force pulling
at one end and fixing the other end of the dsDNA. Single 
molecule experimental studies of DNA elasticity are 
explained well by worm-like-chain model 
\cite{kratkyporod1949,marko1995} which assumes 
inextensibility, isotropic bending rigidity of polymer in 
the thermodynamic limit ($L/L_p \rightarrow \infty$; where $L_p$ is 
the persistence length) \cite{bustamante1994,marko1995}.  
WLC theory gives the average end-to-end distance of the 
polymer when stretched with a force that involves the 
initial contour length and the persistence length as 
fitting parameters. However, the WLC model fails to 
explain the force-extension behavior in the large force 
limit and also for short length of polymers. For example, 
for short DNA molecules the WLC model is inadequate to 
explain the elastic behavior and gives incorrect 
estimate of persistence length \cite{seol2007,wiggins}, 
an intrinsic property of the polymer that is expected to be 
independent of the contour length. In a recent study 
\cite{mathewfenn2008}, it has been shown that shorter 
DNA is softer than measured by single-molecule experiments. 
It was also shown that the variance in end-to-end distance 
has a quadratic dependence on the number of base-pairs 
rather than a linear dependence, a result of linear elastic 
rod model \cite{wiggins}. These failures of the WLC are 
mainly due to finite length effects, boundary conditions 
and rotational fluctuations at the force attachment. Some of 
these corrections have been incorporated into the WLC model
and led to a more general model called FWLC (Finite WLC) in ref 
\cite{nelsonpc2006}. The finite WLC model is able to predict 
force-extension for a wide range of forces for
polymers with lengths ranging from less than the 
persistence length to infinite chain limit \cite{nelsonpc2006}. 
It can also include the effect of formation of a single 
permanent kink in the polymer. Several researchers have 
studied the force-extension behavior of polymers, single and double stranded DNA with 
improvements to the standard WLC model \cite{bouchiat1999,roland2003,
storm_nelsonpc2003,nelsonpc2006,seol2007,rosa2005,thirumalai2010}
but a complete understanding of the elastic behavior at 
various length scales is not yet well established.\\
 
In this paper, we use numerical simulations to solve the
WLC model and obtain the end-to-end distance distribution
as done earlier \cite{samuelsinha2002}. Supported by the WLC numerical
solution, we demonstrate that the full atomic description of 
dsDNA can give more insight into the elasticity 
at short length scales and how the elastic
properties of short dsDNA change when binding to a dendrimer.
From the equilibrium {contour length} and bending angle
distributions of 38 base-pair dsDNA, we calculate the
stretch modulus and bending persistence length of dsDNA. 
The variance of the end-to-end distance has 
a nearly quartic dependence on the number of base-pairs of dsDNA
which has its origin in bending fluctuations.
By stretching the bare dsDNA in
solvent, we calculate the force-extension curves. 
The stretch modulus calculated from
zero and finite-force methods is in good
agreement with experiments.\\

\section{Methods}
{All atom molecular dynamics simulations of DNA in salt 
solution were carried out in equilibrium as well as in
non-equilibrium.} 
The sequence of 12 base-pair DNA used in our simulation
is d(CGC GAA TTC GCG)$_2$, and that for 38 base-pair 
DNA is d(GCC GCG AGG TGT CAG GGA TTG CAG CCA GCA TCT CGT 
CG)$_2$ and was taken from our earlier works 
\cite{maiti2004,maiti2006,santoshjpcm,santoshbj}. 
To study the effect of dendrimer binding on the 
elasticity of DNA, we have used the G3 PAMAM dendrimer and 38 
base-pair dsDNA complex at neutral pH as reported earlier
\cite{nandy2011}. In equilibrium, 38 base-pairs dsDNA and dendrimer bound 38 base-pairs dsDNA
were simulated separately in explicit solvent. We use  {ff03
force field parameters} of Duan {\it et. al.} 
\cite{ff03} to describe the bonded 
and non-bonded interactions for DNA and the TIP3P model 
\cite{jorgensen} for water. We have used the DREIDING force 
field \cite{mayo1990} to describe the intermolecular interaction 
of the dendrimer. The box dimensions were chosen in order to 
ensure a minimum of 10 \AA~ solvation shell around the 
DNA structure during all simulations. The bare DNA system 
is neutralized with Na$^+$ 
counterions and dendrimer bound DNA is neutralized with 
Na$^+$ as well as Cl$^-$ counterions to account for the
negative charge on DNA and positive charge on dendrimer.
Total system size for equilibrium simulations is 34783
atoms for bare DNA and 179234 atoms for dendrimer bound 
DNA including water and counterions.
{For non-equilibrium stretching of 12 base-pair DNA and
38 base-pair DNA, both strands of one
end of dsDNA were pulled with an external force which increased linearly with time.
The other end of the dsDNA was held fixed. During pulling, we measure the extension of the dsDNA as a function
of the applied force.}
For the stretching simulations, we have added extra water 
along the pulling direction to ensure solvation
of DNA even in fully stretched condition. With this, 
the system size of 38 base-pair bare DNA increases to 
97326 atoms. The total number of atoms including water 
and counterions for the stretching simulation of 12 
base-pair DNA is 27858.
The system energy was minimized by 1000 steps of steepest 
descent minimization followed by 2000 steps of conjugate 
gradient minimization. Translational center-of-mass motions 
were removed after every 1000 steps. NPT-MD was used to get 
the correct solvent density corresponding to experimental 
condition. The long range electrostatic interactions were 
calculated with the Particle Mesh Ewald (PME) method \cite{darden}.
A real space cut off of 9 \AA~was used both for the 
long range electrostatic and short range van der Waals 
interactions. We have used periodic boundary conditions 
in all three directions during the simulation.
During the simulation, bond lengths involving bonds to hydrogen
atoms were constrained using SHAKE algorithm \cite{ryckaert}.
For the equilibrium simulation we have simulated the bare dsDNA for
85 ns and dendrimer bound dsDNA was simulated for 70 ns.
For the stretching of {ds}DNA, we {continue
the simulation} until we get a fully stretched {dsDNA.
The time scale of the simulation at which we get fully stretched DNA
depends on the rate of pulling}.
For 12 base-pair DNA we use a pulling rate 
of 10$^{-5}$ pN/fs which requires about 40 ns and 
for 38 base-pair DNA stretching we use 10$^{-4}$ pN/fs 
which requires 10 ns to get the DNA in the fully
stretched form.\\

From the MD trajectories of both the bare DNA and dendrimer bound DNA simulation,
we have calculated the helix axis, end-to-end distance 
and {contour} length using Curves algorithm developed by 
Skelnar and Lavery\cite{curves1988}. All of these parameters are 
calculated as a function of each base-pair step $n$. Using 
these parameters we have analyzed the {contour length}
distribution $P(L)$, end-to-end distance distribution $P(R)$, 
bending angle distribution
$P(\theta)$, variance of end-to-end distance 
$\sigma_{n}^{2}$ and compared them with those obtained from
WLC model. The WLC model is solved numerically
to get $P(R)$ and force-extension curves
for polymers of any length ranging from highly flexible
($L \gg L_p$) to highly stiff ($L \ll L_p$) polymers. 
Force-extension curves were also obtained from MD 
simulations with external force.\\

\section{Results and Discussion}
\subsection{Equilibrium properties of dsDNA}
\subsubsection{{Contour length} distribution $P(L)$}
\begin{figure*}
\centering
        \subfigure[]
        {
        \begin{overpic}[height=47mm]{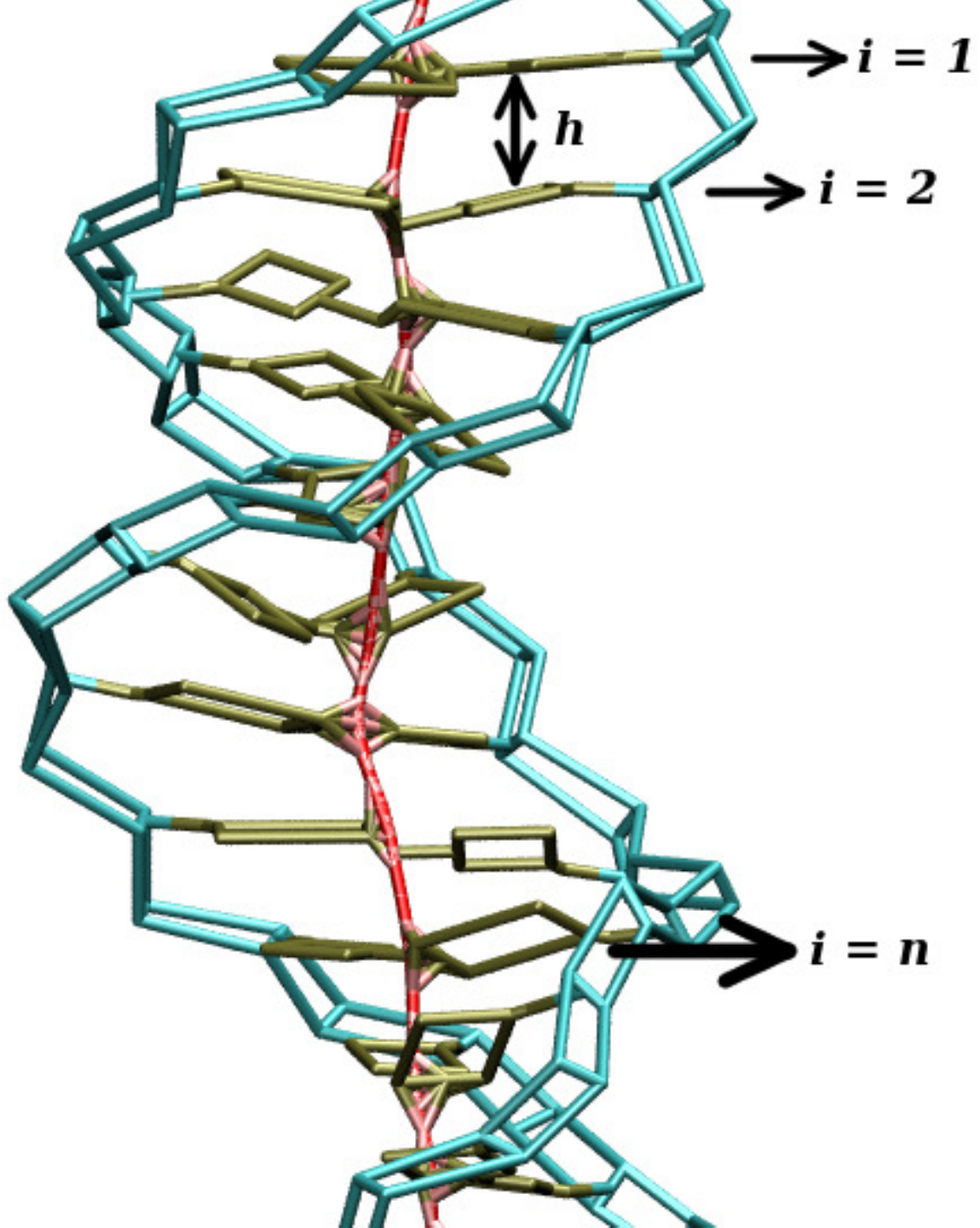}
        \put(41,54){$L = \sum_{i=0}^{n}h_i$}
        \end{overpic}
        \label{schematic_rise}
        }
	\hspace{3mm}
	\subfigure[]
	{
	\includegraphics[height=47mm]{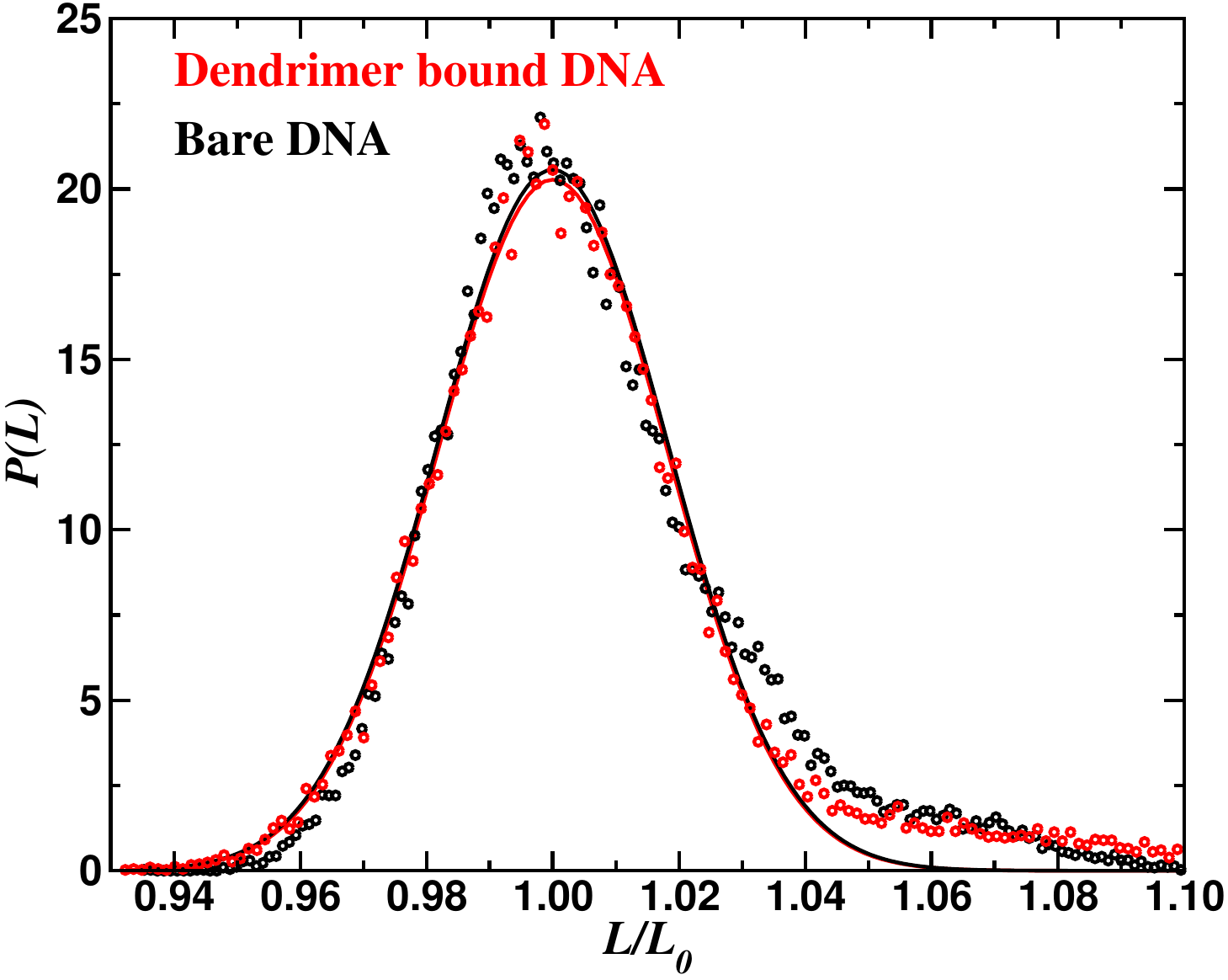}
	\label{pathl_38bp}
	}
	\subfigure[]
	{
	\includegraphics[height=47mm]{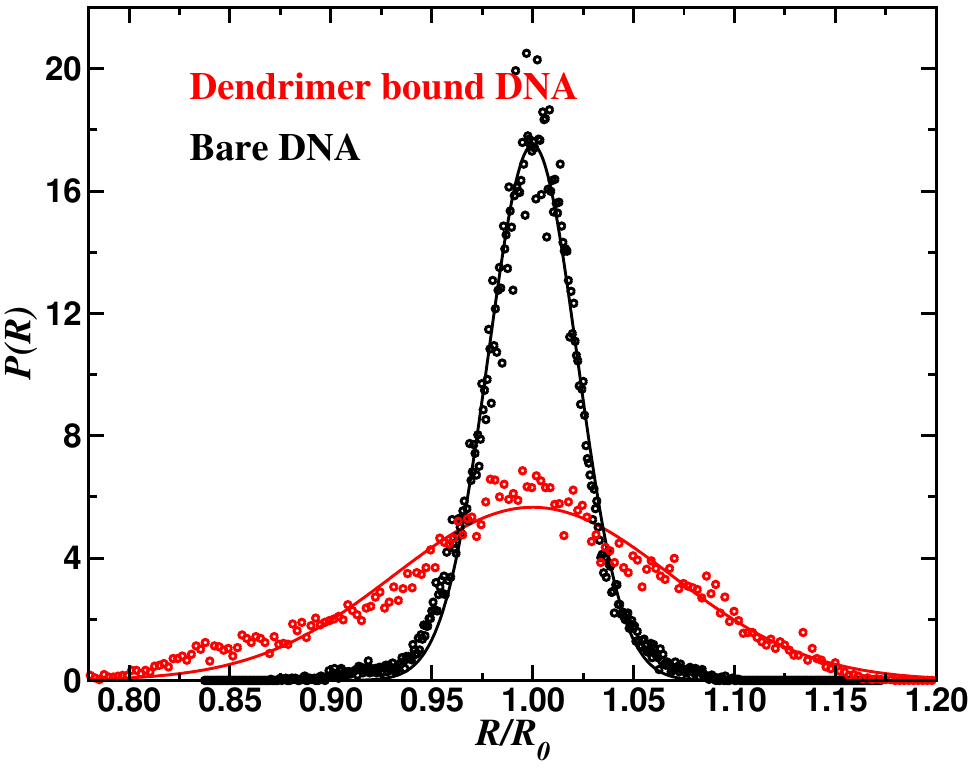}
	\label{pofl_38bp}
	}
	\subfigure[]
	{
	\includegraphics[height=47mm]{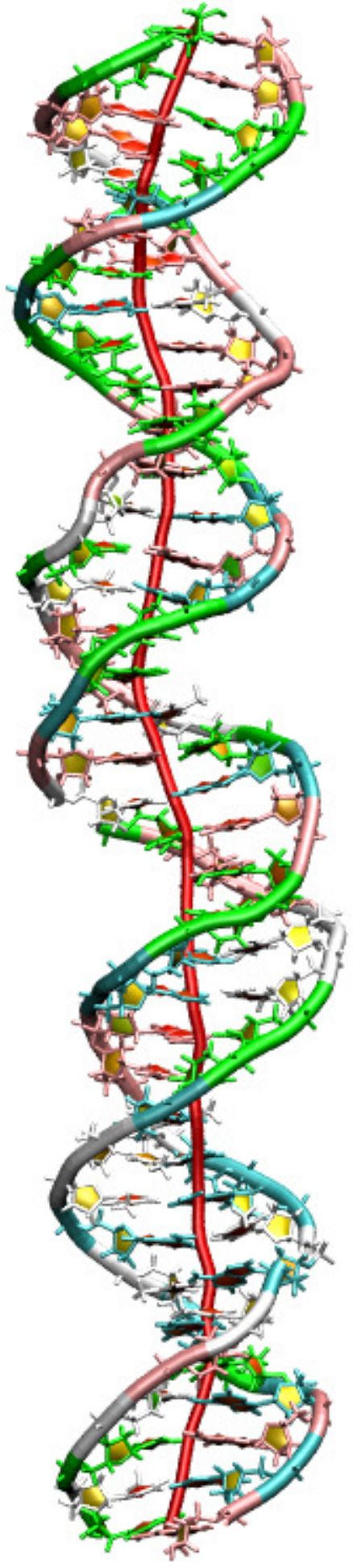}
	\label{38bpdna}
	}
	\subfigure[]
	{
	\includegraphics[height=42mm]{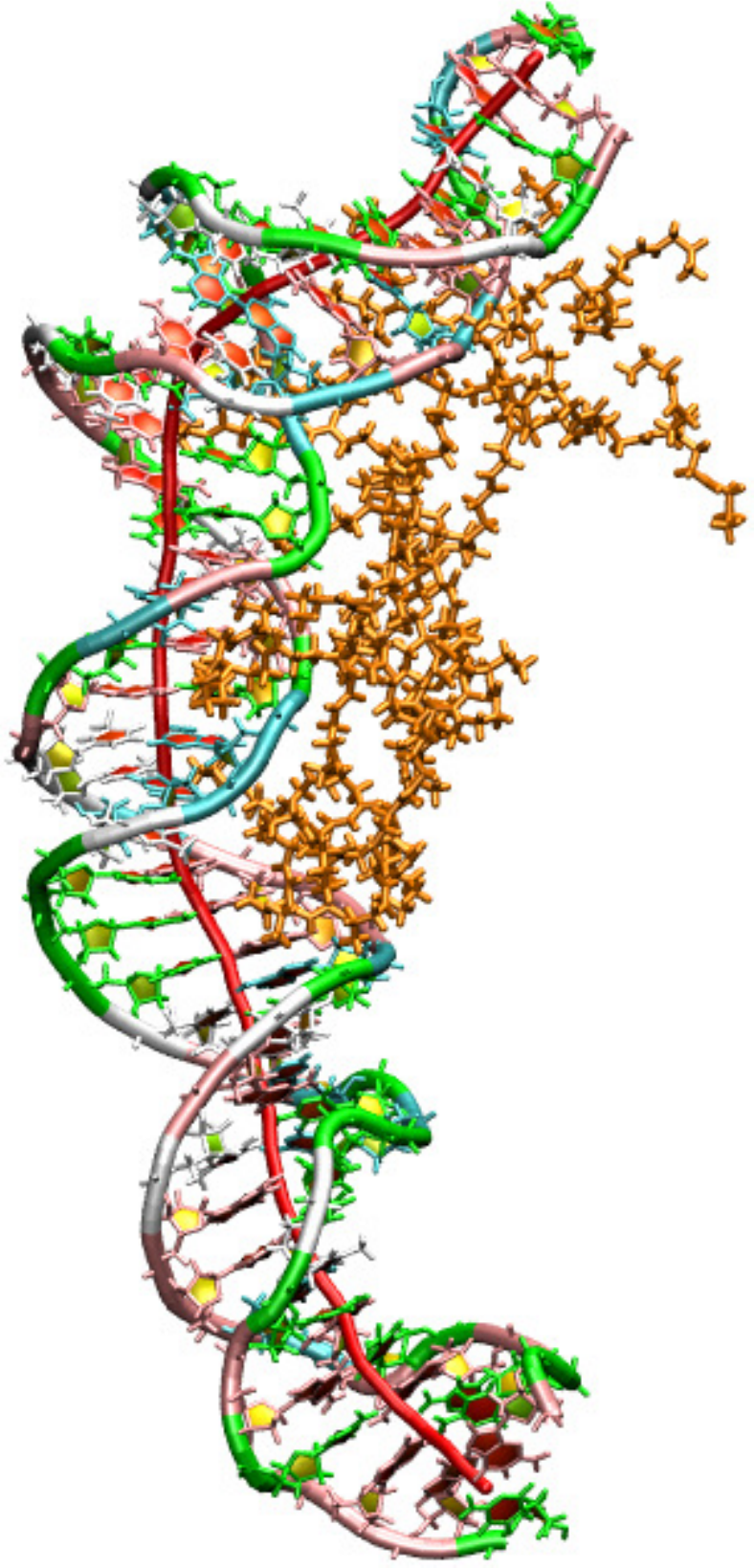}
	\label{g3-24ns}
	}
\caption{\it Equilibrium MD simulations: (a) Schematic showing the base-pair
rise $h$ and contour length $L$. {(b) Contour length and 
(c) end-to-end distance distribution for 
38 base-pair bare DNA and dendrimer bound DNA}. Fitting $P(L)$ to 
a Gaussian gives a stretch modulus $\gamma_1$ of 
{955} pN for bare DNA and {959} pN for dendrimer bound DNA.
Representative snapshots of 38 base-pair (d) bare DNA and
(e) dendrimer bound DNA. 
{We can see from plot of $P(R)$ and snapshots that the bare DNA is 
almost straight with less bending fluctuations, where as dendrimer
bound DNA has large bending fluctuations.}
Color code: adenine - cyan,
guanine - pink, thymine - white, cytosine - green,
helix axis - red and dendrimer - orange.}
\label{poflsnapshots}
\end{figure*}

In equilibrium at room temperature, {the instantaneous 
contour length} ($L$) of dsDNA has thermal fluctuations 
around its mean {contour length} $L_0$. 
{The instantaneous contour length, $L$ is 
defined as the sum of all $n$ base-pair rises, 
$L = \sum_{i=0}^{n}h_i$, where $h_i$ is $i^{th}$ base-pair 
rise as shown in Figure \ref{schematic_rise}.}
A small instantaneous fluctuation ($L-L_0$) 
in {contour length} around its mean value $L_0$ 
generates a restoring force $F$ in the dsDNA that 
is proportional to $L-L_0$, such that $F = -\gamma_1\left(L-L_0\right)/L_0$,
where $\gamma_1$ is the stretch modulus of dsDNA. 
The free energy due to this restoring force can be 
obtained by integrating the force $F$ with respect to 
{contour length}, $E(L) = \frac{\gamma_1}{2L_0}\left(L-L_0\right)^2$. 
Plugging $E(L)$ into the Boltzmann factor {$e^{-\beta E(L)}$, for obtaining the 
probability of having a length $L$ with energy $E(L)$ and normalizing gives
\begin{align}
P(L) &= \sqrt{\frac{\gamma_1 L_0 }{2\pi k_BT}}~e^{-\frac{\gamma_1 L_0}{2k_BT}\left(L/L_0-1\right)^2} \\
\implies \ln~P(L) &= -\frac{\gamma_1 L_0}{2 k_BT}(L/L_0-1)^2 + C
\end{align}}
We have analyzed equilibrium simulation trajectories for 
studying fluctuations in the {contour length} 
of the dsDNA. {Contour length distributions $P(L)$ are shown
in Figure \ref{pathl_38bp} for bare and dendrimer bound DNA.
The contour length distribution for bare and  dendrimer bound DNA is very sharp with a small width.
It means that DNA is stiff with small variance in contour length. 
By fitting $P(L)$ to a Gaussian, we 
obtain the stretch modulus $\gamma_1$ to be 955 pN for bare 
DNA and 959 pN for dendrimer bound DNA.  
The calculated value of the stretch modulus for bare dsDNA is in good agreement with 
experimental reports \cite{bustamante1992,bustamante1994,bustamante3} 
and simulations \cite{santoshjpcm,santoshbj}.
We can see that the dendrimer bound DNA is significantly bent 
around dendrimer compared to bare DNA which is almost straight.
But the bending of DNA around dendrimer does not alter the stretch modulus
since the contour length is almost independent of the degree of DNA bending.
We have also calculated
the end-to-end distance distribution, $P(R)$ to estimate
the degree of bending of DNA around dendrimer.
$P(R)$ and snapshots of bare 
DNA and dendrimer bound DNA are shown in Figures \ref{pofl_38bp}, 
\ref{38bpdna} and \ref{g3-24ns}, respectively.
Note that 
due to the bending of dsDNA, the width of the end-to-end distribution of dendrimer bound DNA is 
very large compared to the width of the distribution for the bare DNA.}\\

\begin{figure*}
	\subfigure[]
        {
        \includegraphics[height=62mm]{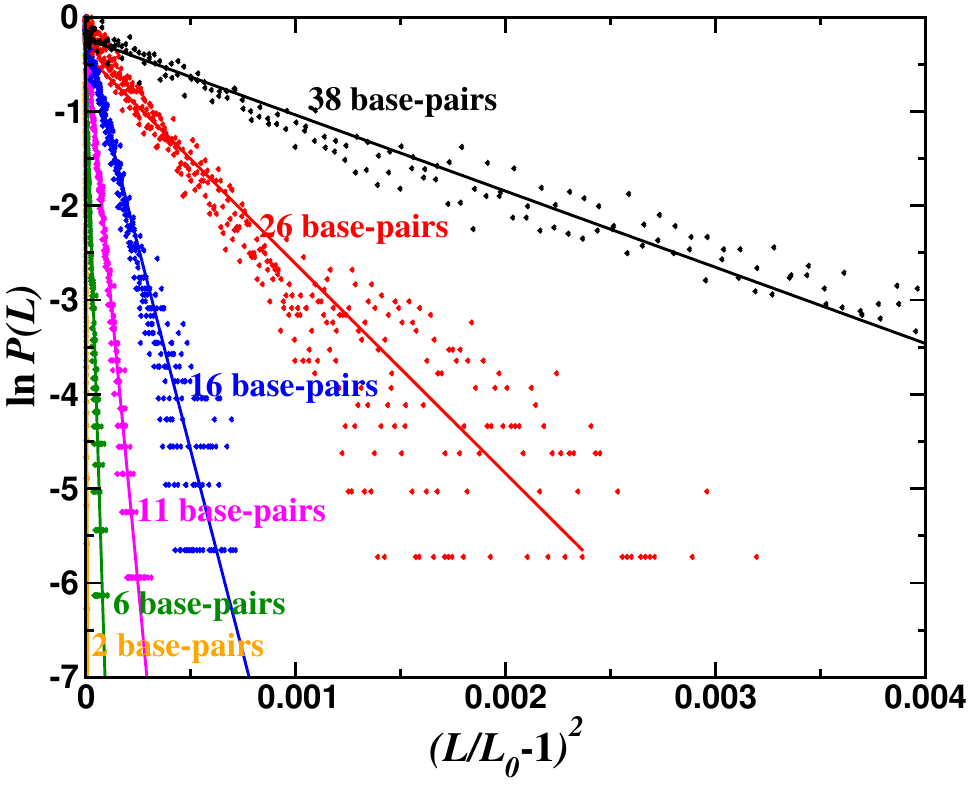}
        \label{logP_lsquare}
        }
        \subfigure[]
        {
        \includegraphics[height=60mm]{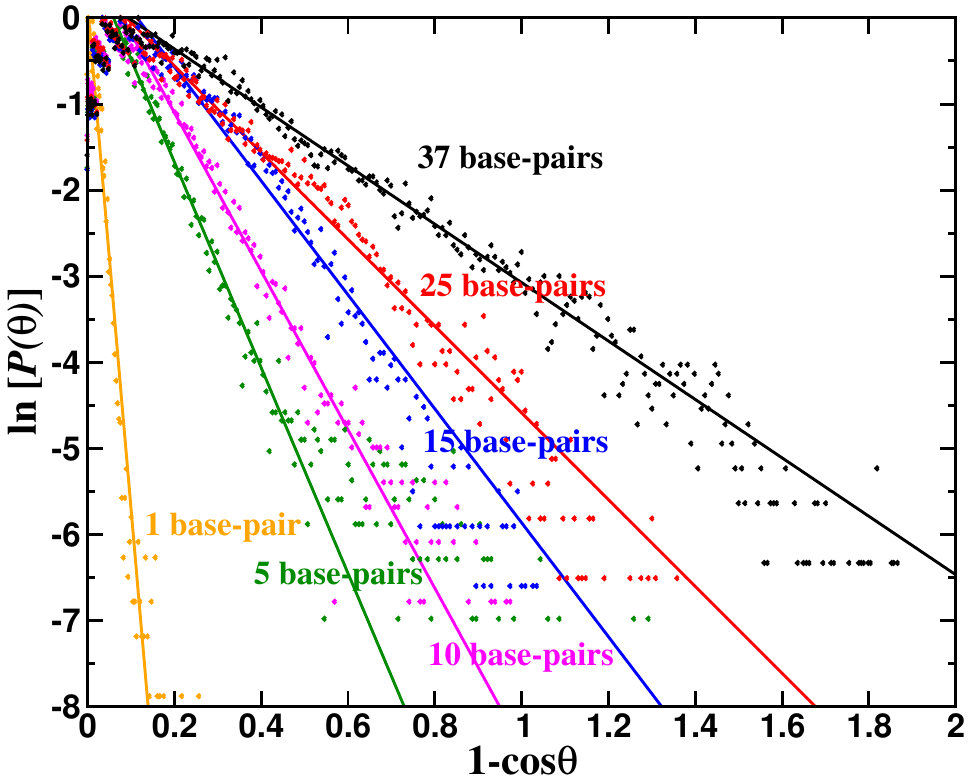}
        \label{correlation_function}
        }
\caption{\it Results for bare DNA: (a) Semi-log plot 
of contour length distribution $P(L)$ for various
base-pair lengths. For all base-pair lengths,
$P(L)$ is Gaussian and hence $\ln~P(L)$ 
is linear in $(L/L_0-1)^2$. (b) Semi-log 
plot of bending angle distribution $P(\theta)$ for various base-pair lengths.
The persistence length $L_p$ calculated from the slope 
for 37 base-pair is 43 nm which is close to experimental findings.}
\label{semilogpoflpoftheta}
\end{figure*}

To calculate the bending persistence length $L_p$
as well as the bending modulus $\kappa$, we calculate
the distribution of bending angle $P(\theta)$. The bending 
angle $\theta$ is defined as the angle between
tangents $t(s)$ and $t(s^\prime)$.
Similar to the {contour length} 
fluctuations, small fluctuation in $\theta$ can 
be approximated to be of Gaussian nature and can be written as
\begin{align}
P(\theta) &= \sqrt{\frac{\kappa}{2\pi {\left|s^1-s^n\right|_{av}} k_BT}}e^{-\frac{\kappa}{2 {\left|s^1-s^n\right|_{av}} k_BT}\theta^2}\\
\implies \ln~P(\theta) &= -\frac{L_p}{{\left|s^1-s^n\right|_{av}}}(1-\cos\theta) + C
\end{align}
{where $\left|s^1-s^n\right|_{av}$ is the 
average contour length ($L_0$)}, $\frac{\kappa}{k_BT} = L_p$ and $\kappa$
is the bending modulus of DNA.
From the simulation trajectories of 38 base-pair 
DNA, we have analyzed $P(L)$ and $P(\theta)$
for different base-pair lengths $n$ between
1 to 38 base-pairs as shown in Figures 
\ref{logP_lsquare} and \ref{correlation_function}, 
respectively. $P(L)$ for all 
base-pair lengths is Gaussian (Figure \ref{logP_lsquare}). 
$\ln P(\theta)$ versus $(1-\cos\theta)$ is 
plotted in Figure \ref{correlation_function}.
$\ln P(\theta)$ is linear in $(1-\cos\theta)$
and from the fit, we obtain the persistence length, 
$L_p$ to be 43 nm which is close compared to the 
standard experimental value of 50 nm. The bending modulus
$\kappa$ of bare DNA obtained from our simulation is 1.76
$\times$ 10$^4$ pN \AA$^2$.\\

{Do the stretch modulus and persistence length calculated from our atomistic 
MD simulation conform to the idea that DNA can be treated as an isotropic elastic rod?
To probe this we note that for the isotropic elastic rod model,
the stretch modulus $\gamma_1$ is related to the persistence length as follows:
$L_p = \gamma_1 r^2/4k_BT$,
where $r$ is the radius of DNA. Using $r$ = 1 nm and $L_p$ = 43 nm,
we estimate $\gamma_1$ for bare DNA to be 705 pN as compared to $\gamma_1$ = 955 pN from contour 
length distribution.
Conversely, for an effective radius of $r$=0.86 nm,
the isotropic elastic rod model works perfectly.
For dendrimer bound DNA, we find an effective persistence length
of $L_p$ = 6.3 nm from our analysis of end-to-end distance
distribution, and again using elastic rod formula
we estimate $\gamma_1$ of dendrimer bound DNA to be 103 pN. This is very small
compared to the value of 959 pN obtained from contour length distribution. 
This implies that dendrimer bound DNA is more flexible than
bare DNA.
Dendrimer is a flexible molecule with positive charges on the 
primary amine groups on the periphery. The branches are mobile 
making the positive charges moving along the negative charges.
When positively charged dendrimer binds to 
negatively charged DNA, charge neutralization 
happens reducing the phosphate-phosphate 
repulsion in the DNA backbone and the 
stiffness of DNA is much reduced.
This also causes DNA to bend around dendrimer. 
We expect similar situations to arise when 
proteins bind to DNA which will be the subject 
of future study.}\\

\subsubsection{Correlations in fluctuations of dsDNA base-pairs}
\label{sec_correlation}
\begin{figure*}
\centering
\includegraphics[height=70mm]{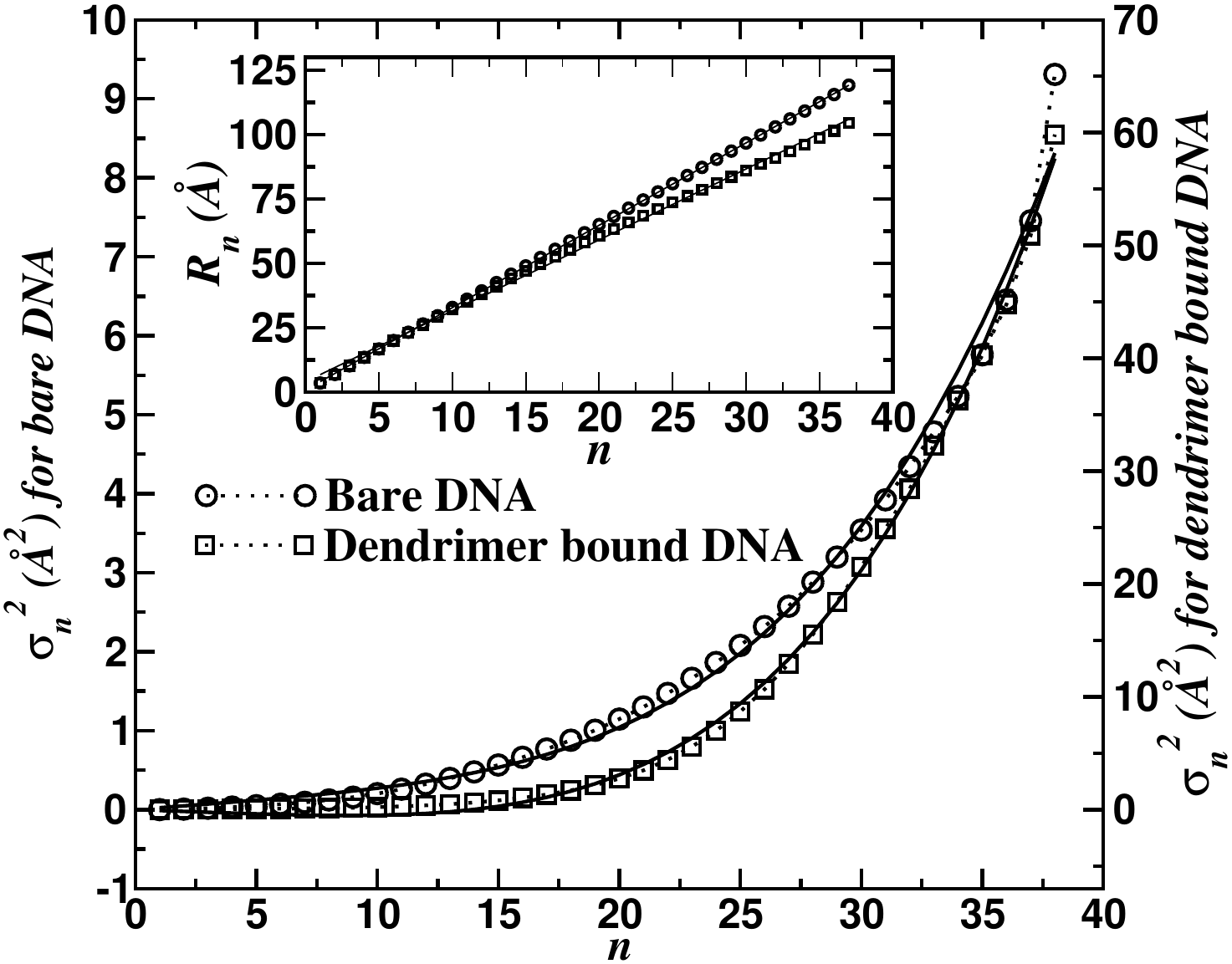}
\caption{\it Variance $\sigma_{n}^2$ in end-to-end distance as a function of
the number of base-pairs $n$ for bare DNA and dendrimer
bound DNA. $\sigma_{n}^2$ is quartic in $n$. Inset shows average 
end-to-end distance $R_n$ as function of $n$, $R_n \propto n$.}
\label{variance_in_l_vs_bp}
\end{figure*}

Mathew-Fenn {\it et. al.} \cite{mathewfenn2008} have
studied DNA flexibility at short length scales
using the variance in the end-to-end distance obtained using
small angle X-ray scattering techniques.
They have tethered clusters of gold atoms to 3$^{\prime}$
thiol linker of DNA ends and measured the distributions
of end-to-end distance for various DNA length of base-pairs 
ranging from 10 to 35. By fitting $P(R)$ to a Gaussian form,
they observe that the variance $\sigma_{n}^{2}$ 
is quadratic with the number of base-pairs $n$ 
in DNA.
Motivated by this study, we also calculate the correlations in 
fluctuations of $P(R)$ for dsDNA base-pairs.
Here, we ask how the room temperature 
fluctuations of individual base-pairs are correlated with 
the neighboring base-pairs in dsDNA by looking at the 
variance $\sigma_{n}^{2}$ in the end-to-end distance, $R$ of dsDNA.
{We write,
\begin{align}
\sigma^2_{n} &= \langle(R_n - R_{n0})^2\rangle
\end{align}
where $R_n$ is the end-to-end distance of $n$ base-pairs and
$R_{n0}$ is its average value.}
We have shown $\sigma^2_{n}$ as a function of 
number of the base-pairs $n$ in Figure
\ref{variance_in_l_vs_bp} for bare DNA and
dendrimer bound DNA, respectively. The inset shows 
end-to-end distance which is proportional to $n$.
We fit the simulation data to $\sigma^2_{n} = an+bn^4$, which
describes the data very well. 
Fitting parameters are $a=0.02359 ~{\text \AA}^2, b=0.00000356 ~{\text \AA}^2$ for
bare DNA and $a=-0.07891~{\text \AA}^2, b=0.0000291~{\text \AA}^2$ for dendrimer bound DNA.
The quartic
term is due to bending fluctuations, while the linear term
accounts for the possible presence of stretching fluctuations.
Our results on $\sigma^2_{n}$ are thus only partially consistent 
with Mathew-Fenn {\it et. al.} \cite{mathewfenn2008}. 
{Here it is worth mentioning 
that this issue of cooperative base-pair fluctuation and its 
relevance in the context of quadratic dependence of the variance of the end-to-end distance 
has been discussed in the literature extensively in last few years 
\cite{mazur2006,mazur2007,mazur2009,mazurpre2009,ranjith2013}. Mazur
attributed this quadratic dependence to the incomplete 
subtraction of the bending contribution from the 
end-to-end distance variance. Becker and Everaers \cite{everaers2009} 
attributed this to subtle linker leverage effect
and concluded that when the linker effect 
is subtracted from the variance data, the dependence 
will be linear. However, recent work by Noy and 
Golestanian \cite{noy2012} shows that indeed the 
quadratic dependence exists even after the bending 
contribution is removed. They attribute this to 
different modes of deformation in the DNA structure.
Our results seem not to be in line with the results 
by Noy and Golestanian \cite{noy2012}. 
More efforts are needed for a complete understanding of this issue.\\

\subsubsection{Worm-like chain model (WLC)}
The worm-like chain (WLC) model (or Kratky-Porod model)
{for the force-extension relation} 
was proposed to explain the elasticity
of polymers \cite{kratkyporod1949,marko1995}.
{Disadvantage is} that analytical results can
be obtained only in the {asymptotic limits}.
By solving WLC model with the help of numerical 
{evaluation}, we 
show that the model can be applied 
to polymers of small $L_0 ~(< L_p)$ to a good approximation. We apply 
this method to study the elasticity of dsDNA. 
The results are then compared to the results
obtained from MD simulations of 38 base-pair
dsDNA whose length is about 0.25 $L_p$. We also
try to connect the effect of a dendrimer binding
on the elasticity of dsDNA.\\

At room temperature,
fluctuations in dsDNA are induced by thermal
agitations. The worm-like chain \index{Worm-like
chain (WLC)} model assumes that the polymer
is a continuous chain with energy given by
\begin{equation}
\label{wlchamiltonian}
\hat{H}=\frac{\kappa}{2}\int\limits_0^{L_0} \left(\frac{d\hat{t}(s)}{ds}\right)^2ds\\
\end{equation}
where $\hat{t}(s)$ is the tangent vector at $s$ 
on the space curve as shown in the schematic in
Figure \ref{sofr}, with $s$ changing from 0 to 
$L_0$, contour length of the polymer and $\kappa$ is the
bending modulus of the polymer. The persistence length
of the polymer is defined as $L_p = \frac{\kappa}{k_BT}$,
where $k_B$ is the Boltzmann constant.\\
\begin{figure*}
        \subfigure[]
        {
        \begin{overpic}[height=60mm]{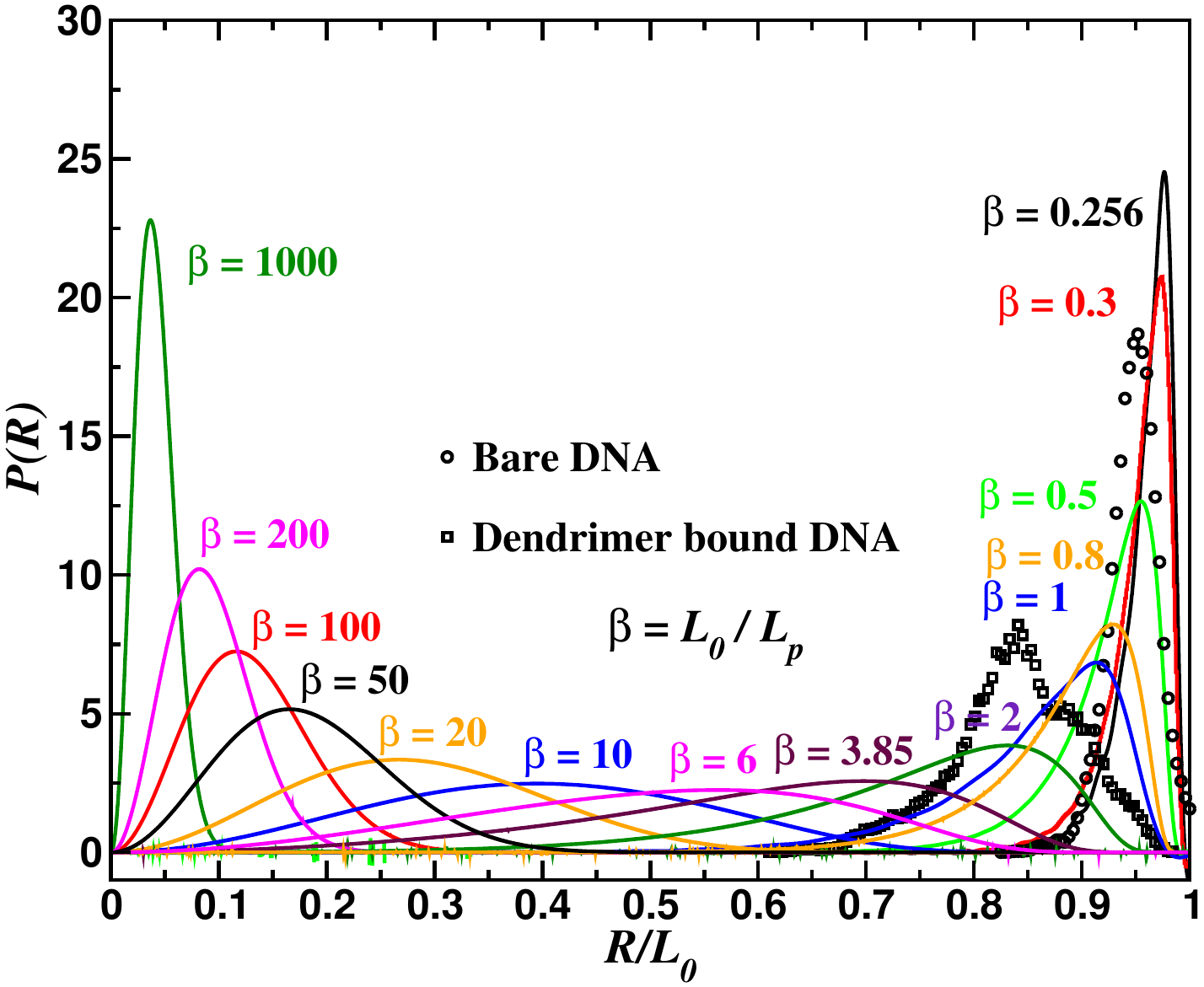}
        \put(25,64){\includegraphics[height=10mm]{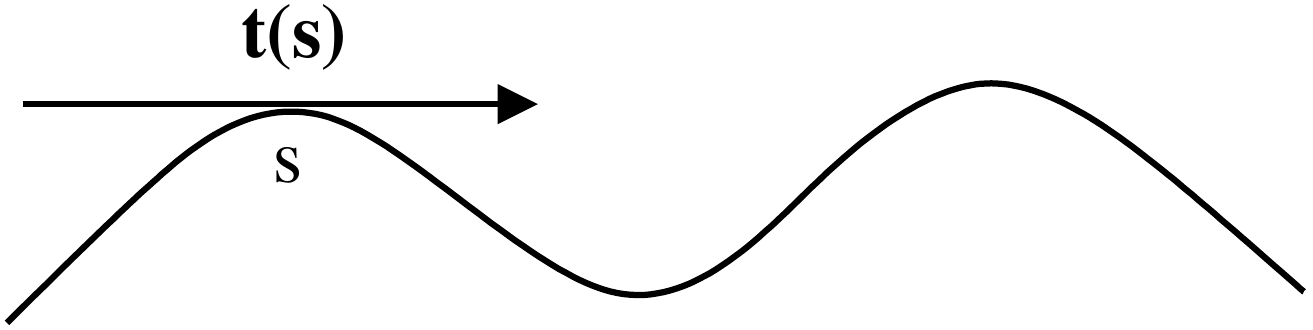}}
	\end{overpic}
        \label{sofr}
        }
        \subfigure[]
        {
        \includegraphics[height=60mm]{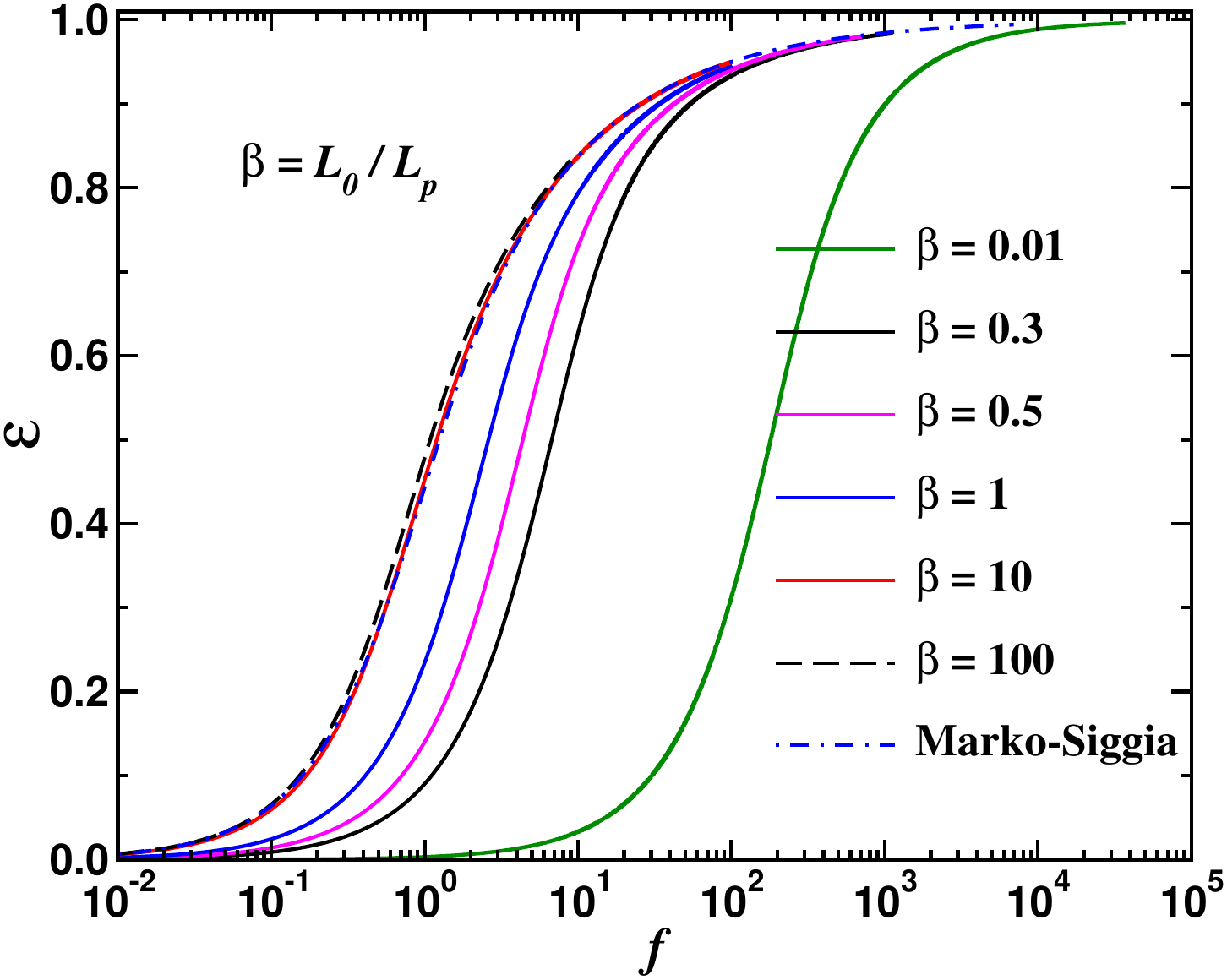}
        \label{fec}
        }
\caption{\it Solution of WLC model using numerical calculations: 
(a) $P(R)$ for various $\beta$ values, where $\beta = L_0/L_p$.
Schematic of the WLC model is shown in the inset.
We compare $P(R)$ of WLC model with $P(R)$ that 
is obtained from MD results for DNA and dendrimer bound 
DNA. $P(R)$ for dendrimer bound DNA is deviating from
the WLC result since WLC model does not consider the 
dendrimer effect.
(b) Force-extension curves for 
various $\beta$ values. Analytical 
results of Marko-Siggia \cite{marko1995}
is in good agreement with force-extension
curve obtained by numerical simulation 
for large $\beta$.}
\label{exactwlc}
\end{figure*}

The numerical solution of the WLC model is described 
in detail by Samuel and Sinha \cite{samuelsinha2002}.
Here we give a brief outline of the method.
Suppose that we apply a force on the WLC polymer at one end 
in the $z$-direction. {The new Hamiltonian can be written as,
\begin{align}
\label{wlchamiltonianforce}
\hat{H}&=\int\limits_0^{L_0}{\left[ \frac{\kappa}{2} \left(\frac{d\hat{t}(s)}{ds}\right)^2-f {t_z}\right]ds}.
\end{align}
Substituting $\kappa=L_p k_BT$ and changing variable to 
$\tau = s/L_p$, we obtain the dimensionless Hamiltonian as,
\begin{align}
\implies \frac{\hat{H}}{k_BT}&=\int\limits_0^{L_0/L_p}{\left[\frac{1}{2} \left(\frac{d\hat{t}(\tau)}{d\tau}\right)^2-\frac{L_p f}{k_BT} {t_z}\right]d\tau}.
\end{align}
Now we use $\beta = L_0/L_p$ and $\tilde f = \frac{L_p f}{k_BT}$ 
to obtain the partition function as,
\begin{equation}
\label{wlcpartitionfunction}
Z(\tilde f)=N\int D[\hat{t}(\tau)]e^{-\int\limits_0^\beta d\tau[\frac{1}{2}(\frac{d \hat{t}}{d \tau})^2 -\tilde f\hat{t}_z]}
\end{equation}
here $\beta = L_0/L_p$ is analogous to the inverse temperature.
Proper choice of the basis set is crucial for the numerical 
evaluation of $Z$. In 3D, the natural choice are the
spherical harmonics $|l\rangle = Y_{l,0} (\theta,\phi)$ 
which are angular part of the normalized eigenfunctions of $-\nabla^2/2$.
To compute $Z(\tilde f)=\langle 0|e^{-\frac{\bf{\nabla}^2}{2}-\tilde f\cos{\theta}}|0\rangle$,
we have used the following basis set to include force $\tilde f$
\begin{eqnarray}
\label{wlcbasisset}
\langle l|\hat{H}|{l^{\prime}}\rangle = \frac{l(l+1)}{2} \delta_{l,l^{\prime}} 
- \frac{\tilde f(l+1)}{\sqrt{(2l+1)(2l+3)}} \left(\delta_{{l+1},l^{\prime}}+\delta_{l^{\prime},{l+1}}\right).
\end{eqnarray}
By calculating $Z(\tilde f)$ numerically to a desired accuracy, 
we can calculate various properties of the system.
The end-to-end distance distribution is 
calculated using, $P(R) = -2R\frac{dp(R)}{dR}$
\cite{samuelsinha2002}, where $p(R)$ is the 
inverse Fourier transform (by working with imaginary $\tilde f$)
of the partition function $Z(\tilde f)$. The resulting expression
for $P(R)$ is given by
\begin{equation}
P(R) = \frac{2R}{\pi}\int\limits_0^{\infty}dk~ k~ Z(ik)~ \sin kR
\end{equation}
where $Z(ik)$ is the partition function under the 
effect of an imaginary force $\tilde f = i k$.
We also calculate the force $\tilde f$ versus extension 
${\epsilon}$ behavior and the free 
energy $G(\tilde f)$, of the system as follows,
\begin{equation}
\label{wlcforceextension}
{\epsilon}=-\frac{d G(\tilde f)}{df}
\end{equation}
and
\begin{equation}
\label{wlcfreeenergy}
G(\tilde f)=-\frac{1}{\beta} \ln Z(\tilde f)
\end{equation}
Where the extension $\epsilon$ is the ratio of
the end-to-end distance to the fixed
contour length of DNA, i.e.,
{$\epsilon = {R}/{L_0}$.}}
These quantities can be measured experimentally giving 
a direct validation of the model.
$P(R)$ and ${\epsilon}$ 
are plotted in Figure \ref{exactwlc}.
$P(R)$ for various values of $\beta$ as a function of
$R/L_0$ is plotted in Figure \ref{sofr}.
For flexible polymers, $\beta \gg 1$ and for stiff
polymers, $\beta \ll 1$ while for semi-flexible polymers 
$\beta$ has intermediate values. In Figure \ref{sofr},
$P(R)$ has a peak at low values of $R/L_0$ for flexible
polymers where as the peak appears at large values of 
$R/L_0$ for semi-flexible or stiff polymers.
This is due to the difference in the energy cost for bending.
The energy cost for bending is less for flexible
polymers compared to semi-flexible polymers. 
Flexible polymers are in coiled form since $L_0\gg L_p$
where as semi-flexible polymers have length comparable 
to $L_p$. We plot $P(R)$ of bare DNA and dendrimer 
bound DNA in Figure \ref{sofr} for comparison.
Bare DNA has peak at $R/L_0 = 0.95$ and dendrimer
bound DNA has peak at $R/L_0 = 0.84$ indicating 
the more flexible nature of DNA when bound to dendrimer.
Comparing the peak position and height, bare DNA has 
persistence length $L_p$ about 41.9 nm which is 
quantitatively in good agreement with the value of 43 nm that was 
calculated from $P(\theta)$ shown in Figure \ref{correlation_function}. 
{By a similar comparison of MD data and the WLC
calculation, dendrimer bound DNA is characterized by
an effective persistence length 
$L_p$ about 6.3 nm implying that the dendrimer
bound DNA is 7 times more flexible than the bare DNA.}
However, the shape of $P(R)$ for dendrimer bound 
DNA is not well correlated with WLC results. This 
is due to the fact that the WLC model does not 
include the interaction
of dendrimer. Moreover in MD simulations we have also
solvent and counterion effects included which are
not included in WLC model.\\

Force-extension curves calculated from WLC model 
are plotted in Figure \ref{fec} and compared with 
the interpolation formula of Marko-Siggia 
\cite{marko1995}. Note that the interpolation formula for force-extension 
curve given by Marko-Siggia \cite{marko1995},
\begin{equation}
\frac{fL_p}{k_BT} = \epsilon + \frac{1}{4\left(1-\epsilon\right)^2} - \frac{1}{4}
\label{eqnmarkosiggia}
\end{equation}
is asymptotically valid for long polymers 
for small and large force limits.
For stiff polymers, the force required to 
stretch is higher compared to flexible polymers.
The Marko-Siggia \cite{marko1995} formula works 
well for large $\beta$ over a wide range of forces.
With the inclusion of the DNA intrinsic 
elasticity via stretch modulus $\gamma_1$,
Odijk \cite{odijk1995}
proposed the following interpolation formula,
\begin{equation}
\label{eqnodijk}
\epsilon = \frac{R}{L_0}=\left[1-\sqrt{\frac{k_B T}{4f L_p}}+ \frac{f}{\gamma_1}\right]
\end{equation}
which is valid for large $\beta$. Eqn. 
\ref{eqnodijk} is used to fit the force-extension
curves obtained from MD simulations which is 
discussed in the next section.\\

\subsection{A note from non-equilibrium stretching}
From the stretching behavior, we can also estimate 
the elastic properties of DNA. For this, we have 
applied external force on one end of the DNA with 
the force applied on the O3$^\prime$ and O5$^\prime$ 
atoms of the two strands of the dsDNA respectively 
while keeping the other end (O5$^\prime$
of one strand and O3$^\prime$ of the other strand)
fixed in order to mimic the single molecule stretching
experiments in atomic force microscopy (AFM) or 
optical or magnetic tweezers. {We have employed
a time varying force ensemble where the force on DNA 
is increased with time and measured the extension
as a function of the momentary pulling force.}
The total energy function of the system 
under the action of external force is given by
\begin{equation}
V({r^N}) = V_o({r^N}) + V_{\text{ext}}(t)
\end{equation}
where $V_o({r^N})$ is the classical empirical
potential describing the bonded and non-bonded 
interactions and $V_{\text{ext}}(t)$ is the potential
under the action of external force
used to stretch DNA which is given by{ 
\begin{align}
V_{\text{ext}}(t) &= (R(0)-R(t)) f(t)
\end{align}
In the above equation, $R(0)=R_0$ is the end-to-end vector distance at time 0,
$R(t)$ is the end-to-end vector distance at time $t$ and $f(t)$ is the time dependent force that acts along
end-to-end vector.} We have used
the force rates of $10^{-4}$ pN/fs and $10^{-5}$ pN/fs
to stretch DNA of 38 base-pairs and 12 base-pairs, respectively
since computational cost increases with the rate of forcing and system size.
However, the obtained force-extension curve strongly 
depends on the rate of force applied.\\

Earlier we have studied the pulling
rate dependence of the dsDNA stretching 
\cite{santoshjpcm,santoshbj}. Force-extension
curves for 12 base-pair DNA at $10^{-5}$ pN/fs 
and 38 base-pair DNA at $10^{-4}$ pN/fs DNA are shown
in Figure \ref{noneq_forceext}. With higher pulling 
rate the plateau in force-extension curve was 
observed at higher pulling force as expected.
\begin{figure*}
        \includegraphics[height=70mm]{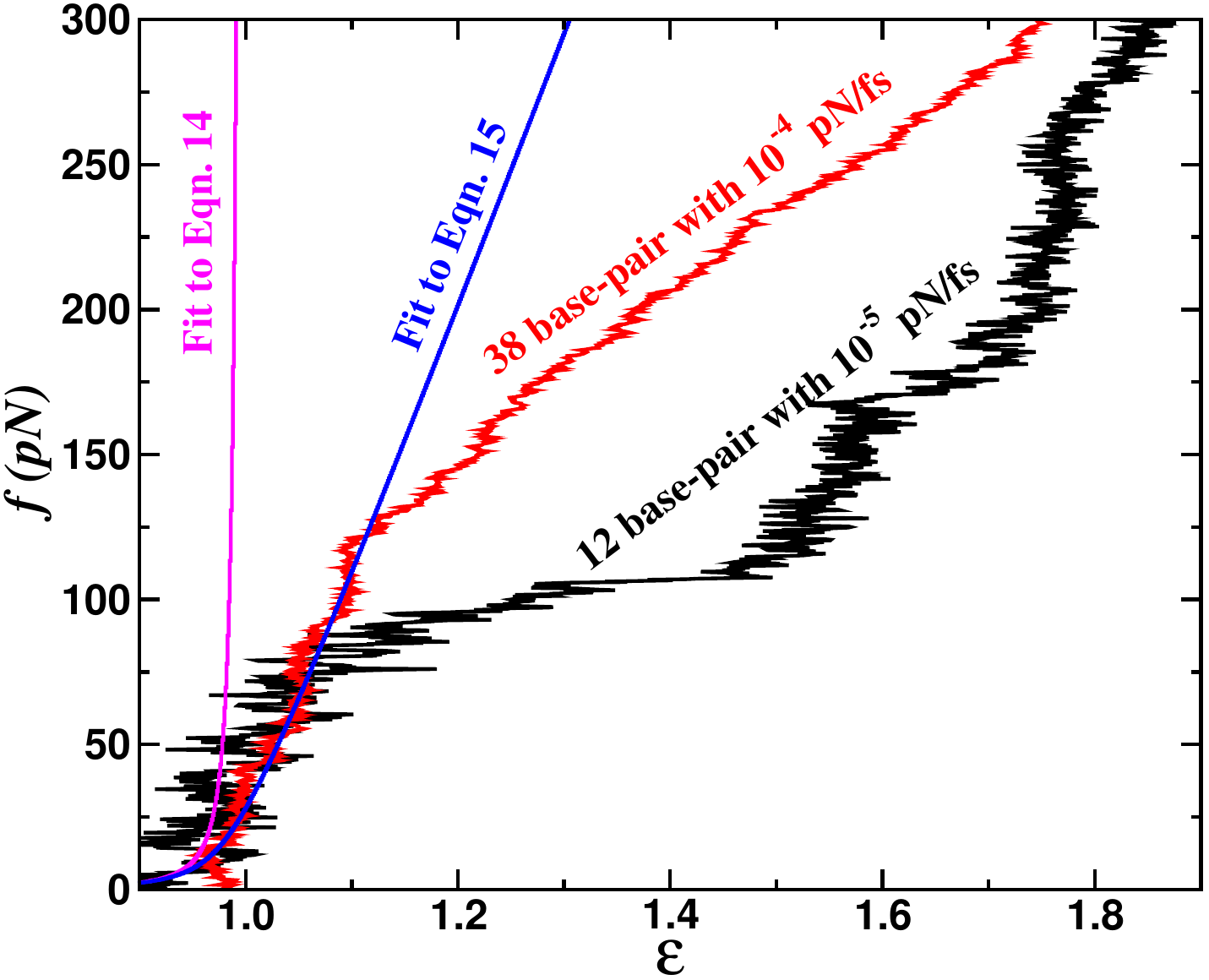}
\caption{\it Non-equilibrium stretching of DNA using MD: 
Force-extension curve for 12 base-pair 
dsDNA at pulling rate of $10^{-5}$ pN/fs and 38 base-pair
dsDNA at pulling rate of $10^{-4}$ pN/fs.
{We fit the force-extension curves 
obtained from MD simulations to 
Eqns. \eqref{eqnmarkosiggia} and \eqref{eqnodijk} 
with $L_0$ as fitting parameter and using
$\gamma_1$ = 955 pN and $L_p$ = 43 nm 
which are obtained from equilibrium $P(L)$
and $P(\theta)$, respectively.}}
\label{noneq_forceext}
\end{figure*}
{The end-to-end distance $R$ is measured as the
average end-to-end O3$^\prime$-O5$^\prime$ atom distance of two strands when 
the force is zero.}
In the initial stage, as the force
is increasing, the length of DNA increases linearly
with the force \cite{santoshjpcm,santoshbj}.
This is followed by a highly nonlinear 
regime called overstretching region where 
DNA gets stretched suddenly about 1.7 times its initial 
length with a very small increment in applied 
force. Our goal here is to understand how well the 
theoretical framework we have discussed so far
helps in understanding the force-extension curves of 
DNA. {We use $\gamma_1$ = 955 pN and $L_p$ = 43 nm
calculated from the equilibrium contour length and angle
distributions to fit force-extension curves with Eqns.
\ref{eqnmarkosiggia} and \ref{eqnodijk}.
Fitting is done with {contour} length $L_0$
as the fitting parameter. In Figure \ref{noneq_forceext}
we fit the force-extension curves to Eqns.
\ref{eqnmarkosiggia}, \ref{eqnodijk} and obtain
contour length $L_0$ of 12.98 nm and 12.84, respectively for
38 base-pair DNA. Similarly fitting of Eqns. \ref{eqnmarkosiggia} 
and \ref{eqnodijk} gives 4.28 nm and 4.3 nm, respectively for 12 base-pair DNA.
It is observed that the enthalpic elasticity 
included in Eqn. \ref{eqnodijk} by introducing $f/\gamma_1$ describes
the simulation data very well for small forces since
the contour length of DNA is much less than $L_p$.}
All of these formulas describe the
force-extension curves quite well for small
forces before the overstretching region. 
But they fail to explain the force-extension
curve in overstretching region. 
The force response in the overstretching transition 
can be fitted by models that include the cooperative 
base-stretching transition \cite{einert2010}, 
which however is not pursued in this paper.
From this analysis, we get good agreement 
between Eqns.\ref{eqnmarkosiggia}, \ref{eqnodijk} and our MD results
for small forces till the overstretching region.
Hence we conclude that the non-equilibrium
force-extension curves also support the observations 
made from equilibrium fluctuation analysis. 
Note that the equilibrium simulations are 
performed at Na$^+$ concentration of 
275 mM for bare DNA and non-equilibrium 
simulations are performed at Na$^+$ concentration 
of 130 mM. Since Eqns.\ref{eqnmarkosiggia} and \ref{eqnodijk}
ignore counterions and solvent effects which are properly
treated in our MD simulations, there could 
be minor mismatch in the fits shown in Figure 
\ref{noneq_forceext}.\\

\section{Conclusion}
We have calculated various elastic properties 
of a dsDNA from equilibrium fluctuation analysis
and numerical solution to WLC model. The distribution of 
equilibrium {contour length} $P(L)$ and bending angle
$P(\theta)$ are fit to Gaussian to calculate the stretch 
modulus $\gamma_1$ and persistence length $L_p$
of DNA, respectively. The stretch modulus obtained from this
equilibrium distribution is compared to the value obtained 
from the non-equilibrium
stretching simulations. We also study how the elasticity
of DNA is affected by protein binding with DNA, considering
dendrimer as a model protein. We find that the
DNA becomes 7-8 times {flexible with
respect to the fluctuations in the end-to-end distance} when dendrimer
binds to it. In the presence of dendrimer, the stretch 
modulus $\gamma_1$ of the DNA is {959} pN and the effective persistence
length $L_p$ is 6.3 nm compared to $\gamma_1$ = {955} pN and 
$L_p$ = 43 nm for bare DNA. 
The calculated elastic parameters are 
in good agreement with the experimental calculations. 
Further by performing numerical calculations to solve 
WLC model, we calculate end-to-end distance distribution 
and force-extension curves of DNA for various lengths 
ranging from highly flexible to highly stiff polymers. 
We find good correlation with the equilibrium results. 
Force-extension curves for 12 base-pair DNA at 10$^{-5}$ pN/fs 
and 38 base-pair DNA at 10$^{-4}$ pN/fs were obtained. The non-equilibrium stretching
simulations are compatible with the results
obtained from equilibrium simulations. These
results are helpful in understanding DNA 
elasticity at small length scales and the effect
of protein interaction with DNA which is abundant
in many cellular phenomena.\\

\section{Acknowledgments}
We thank Yann von Hansen, Supurna Sinha and Rob Phillips for helpful discussions.
SM acknowledges Erasmus Mundas fellowship from European Union
and senior research fellowship from India. 
We acknowledge Department of Biotechnology
(DBT), Government of India for financial support.

\end{document}